  \documentclass[prl,aps,twocolumn,showpacs]{revtex4}
\usepackage[dvips]{graphicx}
\usepackage{dcolumn}
\newcommand{\be}{\begin{equation}}
\newcommand{\ee}{\end{equation}}
\newcommand{\bea}{\begin{eqnarray}}
\newcommand{\eea}{\end{eqnarray}}
\newcommand{\nn}{ \nonumber}

\begin{document}
\topmargin=-15mm

\title{ On the Theory of Magnetotransport in a Periodically Modulated
Two-Dimensional Electron Gas}
\author{Natalya A. Zimbovskaya}

\address{Department of Physics, The City College of CUNY, New York, NY,
10021}

 \begin{abstract} 
 A semiclassical theory based on the Boltzmann transport equation for a two-dimensional electron gas modulated along one direction with weak electrostatic or magnetic modulations is proposed. It is shown that oscillations of the magnetoresistivity $ \rho_{||} $ corresponding to the
current driven along the modulation lines observed at moderately low magnetic fields, can be explained as classical geometric resonances reflecting the commensurability of the period of spatial modulations and the cyclotron radius of electrons.
  \end{abstract}

\date{\today}
\maketitle


Theoretical studies of magnetotransport properties of a 
two-dimensional electron gas (2DEG) spatially modulated along a certain direction due to periodic electrostatic or magnetostatic fields are lasting more than a decade. At present the theory of magnetotransport in modulated 2DEG is
well developed and the most of effects observed in such systems at low magnetic fields have been explained both by quantum mechanical (in a semiclassical limit) \cite{one,two,three,four,five,six,seven}  and classical \cite{eight,nine,ten,eleven,twelve} transport calculations, giving consistent results. One of a few exceptions is an effect of oscillations of the resistivity component $ \rho_{||}$ which corresponds
to the current driven in parallel with the lines of modulation. These oscillations were observed along with the commensurability oscillations of another resistivity component $ \rho_\perp $ corresponding to the current driven across the modulation lines, and have the same period and a phase
opposite to those of $ \rho_\perp $. 

The oscillations of $ \rho_{||} $ had been immediately explained as an effect of a pure quantum nature which
originates from the quantum oscillations of the electron density of states (DOS) in the applied magnetic field $\bf B$ \cite{one,three}. However, actual results obtained in these works \cite{one,three} are not consistent with these conclusions. This can be easily demonstrated, basing on the results of Ref.\cite{three} which are reproduced here (Fig.1).  It is clearly seen in Fig.1 that both $\rho_\perp $ and $ \rho_{||} $ exhibit oscillations of the same period at low (unquantizing) magnetic fields where quantum oscillations of DOS are not resolved, and only effects controlled with classical mechanisms could survive.

\begin{figure}[t]
\begin{center}
\includegraphics[width=8cm,height=7cm]{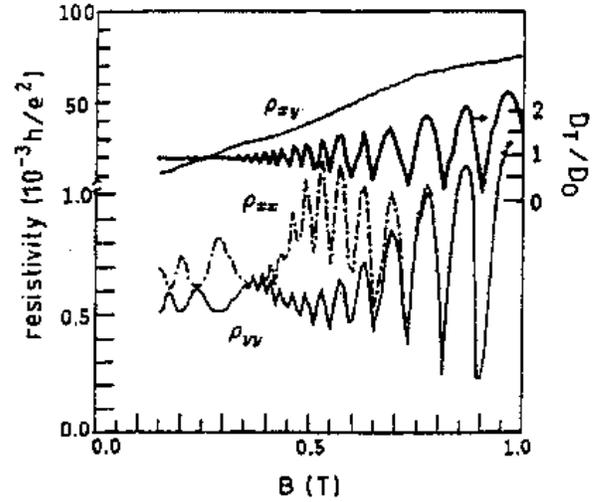}
\caption{
Magnetoresistivity components is magnetic field in a modulated 2DEG with electrostatic modulations applied along the $x$-axis \cite{three}. Within the adopted geometry the component $ \rho_{xx}, \rho_{yy} $ correspond to $ \rho_\perp $ and $ \rho_{||} $ of the present work, respectively. The thik solid line represents the electron DOS in units of $D_0 .$
}  
\label{rateI}
\end{center}
\end{figure}

It is very well known that periods of quantum oscillations and semiclassical commensurability oscillations (geometric resonance) are different. Their ratio equals  $2\pi/\lambda k_F $ where $ \lambda $ is the space period of modulations, and $ k_F $ is the Fermi wave vector of electrons \cite{thirteen}. Therefore, the observed coincidence on periods of the low-field Weiss oscillations of the resistivity component $ \rho_\perp $
and the weaker anti-phase oscillations of $ \rho_{||} $ gives an evidence that for both resistivity components these oscillations are geometric oscillations. Such oscillations arise in modulated 2DEG due to the periodic reproduction of commensurability between the diameters of electron cyclotron orbits and the period of applied  modulations $ \lambda. $ Consequently, we can treat the low-field oscillations of $ \rho_{||} $ as an effect basically following from classical mechanisms, and develop a theory of this effect based on Boltzmann transport equation. The results for the magnetic field dependence of $ \rho_{||} $ given by the classical transport calculations have to be in agreement with the results obtained within the quantum mechanical approach in the low-field limit, as well as in the case of Weiss oscillations of the magnetoresistivity component $ \rho_\perp. $

The existing classical theory of magnetotransport in modulated 2DEG fails to provide  a description of the low-field oscillations of $\rho_{||},$  and concludes that this magnetoresistivity component is not affected by the modulations \cite {fourteen}. This gives a motivation for the present work. Its purpose is to demonstrate that the most important characteristic features of the low-field oscillations of the resistivity $ \rho_{||} $  can be qualitatively reproduced within a semiclassical transport theory. We also briefly analyze where the earlier semiclassical theory might have erred. To simplify the following calculations the effects of anisotropy in electron scattering are neglected, and the relaxation time approximation is used. It is also assumed that the external magnetic field is moderately weak so that the electrons cyclotron radius $ R $ is considerably smaller than their mean free path $ l $ but larger that the period of modulations $ \lambda, $ and $ R >> \sqrt{l \lambda} $ which provides preferred conditions for observation of commensurability oscillations of transport coefficients of the 2DEG. To facilitate a comparison of the present results with the previous work we intentionally use a theoretical formalism as close as possible to that adopted in the current semiclassical theory \cite{eight,nine,ten,eleven}.

We consider first electrostatic modulation with a single harmonic of period $ \lambda =\displaystyle{ 2\pi/g} $ along the $ y $ direction given by $ \Delta E (y) = \displaystyle{- dV(y)/ dy}. $ The screened modulation potential $ V(y) $ is parametrized as $ e V (y) = \varepsilon E_F \sin (gy) $ where $ E_F $ is the Fermi energy of the 2DEG. We examine weak modulations, so that $ |\epsilon g l| << 1. $ Electron transport coefficients for this case were first calculated by Beenakker \cite{eight}. He started from a linearized Boltzmann transport equation for the electron distribution function $ \Phi (y, \psi ) $ of the form
  \be 
 D [\Phi] + C [\Phi] = \bf E \cdot v
  \ee
 where $ \psi $ is the angular coordinate of the electron cyclotron orbit; $ \bf E $ is a homogeneous electric field applied to the 2DEG alongside with the modulating field $ \Delta E (y); $ the collision term $ C [\Phi] $ is written in the relaxation time approximation with the relaxation towards the local equilibrium distribution, namely:
  \be 
 C [\Phi] = \frac{1}{\tau} \left ( \Phi (y,\psi) - \frac{1}{2\pi} \int_0^{2 \pi} \Phi(y,\psi) d \psi \right ),
  \ee
 and the drift term in the case of electrostatic modulations applied along the $ y   $ direction equals
   \be 
 D [\Phi] = v (y) \sin \psi 
\frac{\partial \Phi}{\partial y} + (v'(y) \cos \psi +\Omega )
\frac{\partial \Phi}{\partial \psi} .
  \ee 
   Here, $ \Omega $ is the electron cyclotron frequency and the electron velocity vector $ {\bf v} (y) $ has a direction $ {\bf u} (\psi) = (\cos \psi, \sin \psi), $ and the magnitude $ v (y) = v_F \sqrt{1 + \epsilon \sin (gy)} $ where $ v_F $ is the Fermi velocity in the unmodulated 2DEG. The Eq.(1) with the collision and drift terms of the form (2) and (3) agrees with transport equations of later papers [9--11]. The derivation of this equation is presented in detail in earlier works (see e.g. [10]), and we omit it for brevity.

The electron 
current density in the 2DEG modulated  along the $ y $ direction also depends on $ y $ and can be written as
   \be 
{\bf j} (y) = D_0 e^2 \int_0^{2\pi} \frac{d \psi}{2 \pi} {\bf v} (y, \psi)
\Phi (y, \psi) \, .
           \ee
 Here, $ D_0 = \displaystyle{m/\pi \hbar^2} $ is the electron DOS on their Fermi surface; $ m,\ e $ are the effective mass and the charge of an electron, respectively. 

Following the usual way [10] we write  $ \Phi
(y,\psi) $ as:
    \be 
 \Phi (y,\psi) = \Phi_0 (\psi) \frac{v (y)}{v_F}+ \rho_0 \tau \chi (y,\psi)
  \ee
  where $ \rho_0 $ is Drude resistivity, and $ \tau $ is the relaxation time. The homogeneous distribution function $ \Phi_0 (\psi) = \rho_0 \tau {\bf v}_0 {\bf j}_0 $ describes the linear response of the 2DEG to the field $ \bf E $ in the absence of modulations, and the function $ \chi (y, \psi) $ satisfies the transport equation:
    \be 
D[\chi] + C[\chi] = v'(y) v(y) j_{0y}.
           \ee
 Here, as well as before, $ {\bf j}_0 $ is the current density for the unmodulated 2DEG.

We have to remark that these Eqs.(1) and (6) include some inaccuracies. Namely, the effect of electrostatic modulations is taken into account simply replacing $ v_F $ by $ v(y) $  which is defined above. This is an oversimplification. For systematic consideration we should explicitly introduce into the transport equation an additional term which describes the 'electrochemical' field arising due to inhomogeneity of electron density caused by modulations. As a result the electric field $ \bf E $ has to be replaced by $ {\bf E} - (1/e) \nabla \mu $ where $ \mu (y) $ is the correction to the chemical potential of electrons caused by modulations. The latter provides the local equilibrium of the 2DEG with a stationary electron density $ (\partial n /\partial t = 0 ) . $ The value of this correction $ \mu (y) $ is determined with the continuity equation which in this case has the form $ \nabla \cdot {\bf j} = 0. $

The above inaccuracies do not change the main approximation for the distribution function $ \chi (y, \psi) $ but they can influence next terms in the expansion of $ \chi (y, \psi) $ in powers of a small parameter $ (\Omega \tau)^{-1} $, as we show below. Therefore, using Eq.(6) in further calculations we have to keep only those terms in the expansion of $ \chi (y, \psi) $ which give for the current density results consistent with the continuity equation, and all remaining terms in this expansion have to be ignored.

To proceed we expand $ \chi (y,\psi) $ in a Fourier series in the spatial variable $ y $ which gives the system of differential equations for the Fourier components. 
  \bea 
& {}&\frac{d \chi_0}{d \psi} + \frac{1}{\Omega} C [\chi_0] 
\nn \\
&=& \frac{gR \epsilon}{4} \sin \psi (\chi_1 + \chi_{-1}) -
\frac{gR \epsilon}{4} \cos \psi \left (\frac{d \chi_1}{d \psi} + \frac{d \chi_{-1}}{d \psi} \right ); \nn \\ 
 & \pm & igR \sin \psi \chi_{\pm 1} + \frac{d \chi_{\pm 1}}{d \psi} + \frac{1}{\Omega} C [\chi_{\pm 1}]
  \nn  \\
& = & - \frac{v_F}{4} g R \epsilon j_{0y} + \frac{gR \epsilon}{2} \sin \psi\chi_{\pm 2} 
 \nn \\ 
&-&
 \frac{gR\epsilon}{4} \cos \psi
\left(\frac{d \chi_0}{d \psi} + \frac{d \chi_{\pm 2}}{d \psi} \right ).
    \eea
  The equations determining Fourier components $\chi_{\pm n} (\psi)$ for the $ n \geq 2 $ are given by
   \bea 
 & \pm& gnR \sin \psi \chi_{\pm n} + \frac{d \chi_{\pm n}}{d \psi} + \frac{1}{\Omega} C [\chi_{\pm n}]
  \nn \\ &=&
\frac{gR\epsilon}{4} \sin \psi [n (\chi_{\pm n + 1} - \chi_{\pm n-1} ) + \chi_{\pm n +1} + \chi_{\pm n-1} ]
  \nn \\ 
&=& - \frac{gR\epsilon}{4} \cos \psi \left (
 \frac{d \chi_{\pm n +1}}{d \psi} +
 \frac{d \chi_{\pm n -1}}{d \psi} \right ).
  \eea

Solving these equations, and keeping the terms of the
order or larger than $ (\varepsilon gR)^2 $ we arrive at the following approximation for the distribution function $ \chi (y, \psi):$
   \bea 
&& \chi  (y,\psi)=
 - \frac{v_F}{2} \epsilon g l j_0^y Q \left (\cos \psi - \frac{sin \psi}{\Omega \tau} \right ) 
  \nn \\
& \times & \bigg \{\sin (g R
\cos \psi + gy) - \frac{1}{2} \epsilon gR 
\nn \\ & \times &
\bigg [ \cos (g R \cos \psi) \cos^2
gy - \frac{1}{2} \sin (g R \cos \psi) \sin 2gy \bigg ] \bigg \}
        \nn \\   \eea
 where
 \be 
Q = \frac{J_0 (gR)}{1 - J_0^2(gR)} \,,
           \ee
 and $ J_0 (gR) $ is the Bessel function. Using the obtained distribution function we can easily calculate the electron current density given by Eq.(4).

Taking into account only greatest terms in the expansion of $ \chi (y, \psi) $ in powers of small parameter $ (\Omega \tau)^{-1} $ we obtain that only $ j_x, $ component gets a correction due to the modulations along the $ y $ direction, whereas the component $ j_y $ remains equal to $ j_y^0 $ and does not depend on spatial coordinates. This agrees with the continuity equation $ \nabla \cdot {\bf j} = 0 $ which is necessary to obtain correct results for electron transport coefficients in modulated 2DEG.

 To proceed we need to define effective (averaged over the period of modulations) transport coefficients. Here, we define the effective conductivity tensor $ \sigma_{eff} $ as follows:
  \be 
{\bf j} \equiv <{\bf j} (y)> \equiv \frac{g}{2\pi} \int_0^{2\pi/g} {\bf j}
(y) dy \equiv \sigma_{eff} \bf E \, .
           \ee
 To justify the adopted definition (11) we point out that the expressions for transport coefficients obtained either with quantum mechanical or classical calculations have to be consistent at low magnetic fields. Quantum mechanical calculations of the magnetoresistivity \cite{three,four,five,six,seven} give the expression which passes over to a classical conductivity tensor averaged over the period of modulations. Therefore, the latter is an accurate semiclassical analog of the conductivity calculated within the proper quantum mechanical approach, and our definition of $ \sigma_{eff} $ agrees with that of Ref. \cite{fifteen}. The same definition was used previously by Mirlin and Wolfle \cite{eleven}.

As a result we obtain:
  \be 
\sigma_{eff}^{xx} = \frac{\sigma_0}{1 + (\Omega \tau)^2} + \frac{1}{4}
(\varepsilon gR)^2 \sigma_0 \frac{J_0^2 (gR)}{1- J_0^2(gR)}
           \ee
 where $ \sigma_0 = 1/\rho_0 $ is the Drude conductivity of the electron system. Other components of the effective conductivity are not influenced due to modulations within our approximation. The second term in Eq. (12) arises due to  modulations. This term exhibits oscillations whose origin is explained below.

It is known that in the presence of both electric and magnetic fields centers of electron cyclotron orbits drift in parallel with the cross product $ \bf E \times B. $ The speed of this motion $ v_{d r} $ equals $ E/ B $ which can be easily shown starting from the equations of motion of an electron subject to these fields. This results in one-dimensional diffusion of electrons along the direction of the drift. The latter brings extra contributions to the current density and changes the  conductivity of the system. Here, electrostatic modulation $ \Delta E (y) $ is applied along the $ y $ direction therefore guiding centers are drifting along the $ x $ axis, and the electron diffusion caused by the drift reveals itself as a corection to $\sigma_{xx} $ component of the magnetoconductivity tensor. The second term in Eq.(12) represents this correction.

To estimate $ v_{dr} $ for the inhomogeneous electric field $ \Delta E (y) $ we have to average $ \Delta E (y) $ over the cyclotron orbit. The mean value is mostly determined with contributions from the vicinities of two stationary points at the orbit where electron is moving in parallel with the modulation equipotential lines (along the $ x $ direction), and the field $ \Delta E (y) $ is varying slowly \cite{eight}. Correspondingly, $ v_{dr}^2 $ is enhanced when $ E (y) $ takes on values of the same sign at both stationary points, and reduced otherwise. So the diffusion correction in the conductivity $ \sigma_{eff}^{xx} $ which is proportional to $ \big <v_{dr}^2 \big > $ ehxibits an oscillatory dependence on the magnetic field, an these oscillations are commensurability oscillations due to geometric resonances in the guiding center drift. This physical mechanism was proposed to explain semiclassical Weiss oscillations in the magnetoresistivity $ \rho_\perp $ \cite{eight}. The present result \cite{twelve} shows that the same mechanism controls semiclassical oscillations of $ \rho_{||} $ (see Eqs.(13) ans (14) below).

The effective magnetoresistivity tensor is defined here as the inverse of the effective conductivity introduced by Eq.(11), $\rho_{eff} = \sigma_{eff}^{-1} $. For the current driven across the modulation lines,
the corresponding resistivity is:
  \be 
\rho_\perp = \rho_{yy} = \rho_0 \bigg \{1 + \frac{1}{4} (\epsilon gl)^2 \frac{J_0^2 (gR)}{1 - J_0^2 (gR)} \bigg \} \, .
           \ee
 Assuming that the current flows along the modulation lines we get:
  \be 
\rho_{||} = \rho_{xx} = \rho_0 \bigg \{1 - \frac{1}{4} 
(\epsilon gR)^2 \frac{J_0^2 (gR)}{1 - J_0^2 (gR)} \bigg \} \, .
           \ee
 For moderately weak magnetic fields $(gR>>1) $ the obtained results (13) and (14) describe oscillations of both magnetoresistivity components periodic in the inversed magnetic field magnitude. The oscillations of $\rho_\perp $ and $ \rho_{||} $ have the same period in $ 1/B $ and the opposite
phases which corroborates the experimental results of \cite{one}. The amplitude of the oscillations of $ \rho_{||} $ is considerably smaller than that of $ \rho_\perp,$ and this also agrees with the experiments of Ref. \cite{one}, and with
the results of numerical quantum mechanical calculations of Ref. \cite{three} in the limit of a weak magnetic field (see Fig.1 above). The result for the resistivity $\rho_\perp
$ as well agrees with the corresponding results of Refs. \cite{eight,nine,ten,eleven} obtained within a classical magnetotransport theory.

However, the expression (14) for $ \rho_{||} $ differs from the well known result of the current semiclassical theory. To analyze this discrepancy we now calculate the current density taking into account next terms in the expansion of the distribution function (9) in powers of $ (\Omega \tau)^{-1}. $ Keeping terms no less than $ (\epsilon g R/\Omega \tau)^2 $ we obtain that now all components of the conductivity get grating-induced corrections: 
   \bea 
 \delta \sigma_{xx} &=& \frac{\sigma_0 (\Omega \tau)^2}{1 + (\Omega \tau)^2} \alpha (y);
  \nn \\
 \delta \sigma_{xy} &=& - \delta \sigma_{yx} = \frac{1}{(\Omega \tau)} \delta \sigma_{xx};
  \\
 \delta \sigma_{yy} &=& - \frac{1}{(\Omega \tau)^2} \delta \sigma_{xx} .
 \nn
   \eea
   Here, $ \alpha(y) $ is the  $ y $ dependent factor of the order of $ (\epsilon g R )^2 $. It follows from Eq.(15) that both components of the current density depend on $ y $ and continuity equation is violated. On these grounds we conclude that only main approximation for $ \chi (y, \psi) $ can be used in magnetotransport calculations, since all remaining terms in the expansion of $ \chi (y, \psi) $ include mistake arising due to original inaccuracies described before. This gives an extra justification to our results (13) and (14) for magnetoresistivities.

At the same time the expressions (15) enable us to show the origin of the mistake in the current semiclassical theory of Refs. \cite{eight,nine,ten,eleven}. With some formal transformations of the transport Eq.(6) we can present $ \big< \alpha (y) \big> $ in the form:
    \be 
 \big< \alpha (y) \big> = \frac{1}{2 \pi} \int_0^{2 \pi} 
\big< v (y) \sin \psi G (y, \psi) \big> d \psi
  \ee
  where $ G(y,\psi) $ satisfies the equation:
  \be 
 D [G] + C [G] = - \frac{2 v' (y) v (y)}{v_F^2}.
  \ee

This gives for $ \sigma_{eff} $ and $ \rho_{eff} $ expressions absolutely consistent with the existing semiclassical theory \cite{eight,nine,ten,eleven,twelve}. Hence, these expressions could be obtained starting from the incorrect expression for the distribution function $ \chi (y, \psi ) $ which includes terms influenced due to inaccuracies in the transport Eqs.(1) and (6). This gives grounds to seriously doubt the results of earlier works \cite{eight,nine,ten,eleven,twelve} especially those concerning $ \rho_{||}. $

So, the present analysis shows that simplified transport Eq.(1) can be successfully used in calculations of the main term in the expansions of transport coefficients in powers of $ (\Omega \tau)^{-1}. $ As for next terms in these expansions, we need to modify the transport Eq.(1) (both drift and collision terms) to get them. For that we need to carry out consistent and systematic consideration of effects of internal electrochemical field arising due to grating-induced inhomogeneity of electron density. This is important for redistribution of the electron density at the presence of modulations  provides the local equilibrium of the system \cite{sixteen}.

Considering magnetic modulations we arrive at similar results. Assuming that the magnitude of the magnetic field gets a weak spatially dependent correction $ \Delta B (y) = \Delta B \sin (g y) \ (\Delta B/B << 1) $ we can derive the transport equation for the contribution to the distribution function of electrons  $ \chi (y, \psi) $ which arises due to the modulations:
   \be 
D[\chi] + C[\chi] = \Delta \Omega (y) (v_{0y} j_x^0 - v_{0x} j_y^0)
           \ee
 where $ \Delta \Omega (y) $ is the correction to the cyclotron frequency due to modulations, and the drift term equals:
  \be 
 D[\chi] = v_{0y} \frac{\partial \chi}{\partial y} + 
(\Omega + \Delta \Omega (y)) \frac{\partial \chi}{\partial \psi}.
  \ee
  Writing out the transport Eq.(18) we have taken into account that the magnetic modulations do not change the magnitude of the electron velocity at the Fermi surface, so that $ v(y) = v_F. $ Due to the same reason the first term in the expression (5) now does not depend on spatial coordinates and equals to the electron distribution function for the unmodulated 2DEG.

Solving Eq.(18) and keeping terms no less than $ (\Delta B/B)^2$ in the solution, we calculate the effective conductivity of the 2DEG at the presence of magnetic modulations following the way used before to analyze the magnetoresistivities of the 2DEG modulated by a weak electrostatic field. As a result we obtain that $ \sigma_{xx} $ component of the effective conductivity tensor gets a correction due to the modulation:
   \be 
 \delta \sigma_{xx} = \sigma_0 \left ( \frac{\Delta B}{B} \right )^2 \frac{J_1^2 (g R)}{1 - J_1^2 (g R)}
  \ee
  where $ J_1 (gR) $ is the Bessel function. As in the case of electric modulations this correction originates from a guiding centers  drift. Using Eq.(20) we easily arrive at the expressions for the magnetoresistivities $ \rho_\perp $ and $ \rho_{||} $ for the current driven across and along the modulation lines:
  \bea 
\rho_\perp &=& \rho_0 \bigg \{ 1 + (\Delta \Omega \tau)^2 
\frac{J_1^2 (gR)}{1 - J_0^2 (gR)} \bigg \} \, ;
          \\ \nn \\
\rho_{||} &=& \rho_0 \bigg \{ 1 - \bigg(\Delta \Omega \tau \frac{R}{l}\bigg)^2 
\frac{J_1^2 (gR)}{1 - J_0^2 (gR)} \bigg \}  \, .
           \eea
 These results describe Weiss commensurability oscillations in the limit of weak magnetic fields $ (g R >>1). $ As well as for electrostatic modulations, the expression (22) reports weaker (due to the small factor $ (R/l)^2 ) $ oscillations of $ \rho_{||} $ whose phase is opposite with respect to that of commensurability oscillations of $ \rho_\perp. $

Finally, the novel result of the paper is that it gives a qualitative explanation of the low-field oscillations of the magnetoresistivity component $ \rho_{||} $ in the 2DEG modulated along one direction within a semiclassical approach. It is shown here that these oscillations of $ \rho_{||} $ at low magnetic fields are commensurability oscillations. At low temperatures when the quantum oscillations of the electron DOS at the Fermi surface are resolved, Shubnikov-de Haas oscillations can be superimposed upon the geometric oscillations of the magnetoresistivity. However, this does not change a classical nature of the effect itself.
\vspace{1mm}

{\it  Acknowledgments:}
I thank G.M. Zimbovsky for help with the manuscript.



\begin{thebibliography}{99}
\bibitem{fourteen} It was shown in Ref.\cite{ten} that within a classical approach the oscillations
of $ \rho_{||} $ could arise due to the oscillations of scattering rates of electrons in the modulated 2DEG. However the results of [10] give in-phase commensurability oscillations of $\rho_{\perp} $ and $ \rho_{||}$ at low magnetic fields instead of anti-phase oscillations observed in the
experiments.
\bibitem{sixteen} The method of calculations adopted in the current theory is mostly based on averaging of transport equation multiplied by a velocity component with respect to both $ \psi $ and $ y $ (see e.g. Refs. \cite{ten,eleven}). This procedure nullifies the contribution from the second term of the collision integral (2) which provides the relaxation of the system towards the local equilibrium. As a result, one arrives at the expressions for transport coefficients corresponding to the relaxation of the system towards the total equilibrium which contradicts the continuity equation. This could result in additional inaccuracies in the expressions for the electronic transport coefficients.
\end{thebibliography}
\end{document}